\documentclass[journal]{IEEEtran}
    
\usepackage{url}
\usepackage{graphicx}
\usepackage{caption}  
\usepackage{subcaption}    
\usepackage{color}
\usepackage[table]{xcolor}
\usepackage{hyperref}  
\usepackage{mathtools}
\usepackage{amssymb}  
\usepackage{amsmath}  
\usepackage{amsthm}
\usepackage{algorithm}        
\usepackage{algpseudocode}    
\usepackage{enumerate}  
\usepackage{tabularx}  
\usepackage{multirow}      
\usepackage{setspace}          
\usepackage[noadjust]{cite}                   

\usepackage[keeplastbox]{flushend}
  
\algrenewcomment[1]{/*#1*/}
\algnewcommand{\LineComment}[1]{\State /*#1*/}
  
\makeatletter
\newcommand{\StatexIndent}[1][3]{%
  \setlength\@tempdima{\algorithmicindent}%
  \Statex\hskip\dimexpr#1\@tempdima\relax}
\algdef{S}[IF]{IfNoThen}[1]{\algorithmicif\ #1}%
\makeatother

\begin{document}  

\title{Machine Learning-based Link Fault Identification and Localization in Complex Networks}

\author{Srinikethan~Madapuzi~Srinivasan, Tram~Truong-Huu,~\IEEEmembership{Senior Member,~IEEE} \\ and~Mohan~Gurusamy,~\IEEEmembership{Senior Member,~IEEE}%
  
  \thanks{Manuscript received November 22, 2018; revised February 28, 2019; accepted March 22, 2019. Date of publication March day, 2019; date of current version Month day, 2019. This work was supported by Singapore MoE AcRF Tier 1 Grant, NUS WBS No. R-263-000-C04-112. \textit{(Corresponding author: Tram Truong-Huu.)}}
  
\thanks{The authors are with the Department of
  Electrical and Computer Engineering, National University of
  Singapore. Singapore 117583. (e-mail:  elemss@nus.edu.sg,
  tram.truong-huu@ieee.org, gmohan@ieee.org).} 

\thanks{Copyright (c) 2019 IEEE. Personal use of this material is permitted. However, permission to use this material for any other purposes must be obtained from the IEEE by sending a request to pubs-permissions@ieee.org.}

}  
    
\markboth{IEEE Internet of Things Journal}%
{S. M. Srinivasan~\textit{et al.}: Machine Learning-based Link Fault Identification and Localization in Complex Networks} 

\maketitle    
  
\begin{abstract}
With the proliferation of network devices and rapid development in information technology, networks such as Internet of Things are increasing in size and becoming more complex with heterogeneous wired and wireless links. In such networks, link faults may result in a link disconnection without immediate replacement or a link reconnection, e.g., a wireless node changes its access point. Identifying whether a link disconnection or a link reconnection has occurred and localizing the failed link become a challenging problem. An active probing approach requires a long time to probe the network by sending signaling messages on different paths, thus incurring significant communication delay and overhead. In this paper, we adopt a passive approach and develop a three-stage machine learning-based technique, namely ML-LFIL that identifies and localizes link faults by analyzing the measurements captured from the normal traffic flows, including aggregate flow rate, end-to-end delay and packet loss. ML-LFIL learns the traffic behavior in normal working conditions and different link fault scenarios. We train the learning model using support vector machine, multi-layer perceptron and random forest. We implement ML-LFIL and carry out extensive experiments using Mininet platform. Performance studies show that ML-LFIL achieves high accuracy while requiring much lower fault localization time compared to the active probing approach. 
\end{abstract}    

   
\section{Introduction}     
\label{sec:intro}

The Internet of Things (IoT), comprising billions of intelligent devices with sensing and processing capabilities along with the ability to connect to the Internet through wired or wireless connections, is being increasingly used to monitor and respond to events in real-time. By $2020$, the number of such Internet-connected devices is expected to exceed $50$ billion~\cite{evans:2011}, raising new challenges in network management. For instance, the traffic generated by Internet vehiculars is expected to reach 300000 Exabyte by the year of 2020~\cite{xu2018internet}. The large number of devices and high network complexity lead to a higher chance of link faults. Such link faults may lead to: (i) a link disconnection without an immediate replacement, e.g., link cut and switch ports down, or (ii) a link reconnection, e.g., a wireless node changes its access point due to poor wireless channel quality. A fault occurring on a crucial link, if not recovered in a timely manner, could lead to the disruption of services to customers. It can take hours or days to repair a link fault by using a manual recovery approach based on the combination of ping, trace-route and other functionalities for Ethernet and MPLS~\cite{banerjee:2012}. Thus, it is essential to have an efficient fault management system, which is able to diagnose faults as soon as possible, and quickly recover the network from such faults, e.g., using a proactive fault recovery mechanism~\cite{mohan:2017}.   
  
The accuracy of fault identification and localization depends on the algorithm used and the correctness of the data (i.e., network state) captured from the network, which in turn depends on the frequency of data collection (i.e., data sampling frequency) and the method by which the data is captured. The network state data can be captured by two methods: active and passive measurements. Active measurements using signaling messages (also known as active probing) have been extensively studied in the
literature~\cite{staessens:2011,adrichem:2014,cheng:2016} and is more commonly used in practice. In an active probing approach, several measurement endpoints (MEPs) or measurement intermediate points (MIPs) are deployed in the network, which inject and exchange additional control packets among themselves to identify and localize network failures. Thus, the accuracy of the active probing approach depends on the number of paths probed and the number of messages injected into the network. On one hand, this increases the communication overhead due to additional traffic injected. On the other hand, it increases the latency in fault identification and localization due to the propagation delay of signalling messages on the probed paths. In contrast to active probing methods, passive monitoring does not inject additional traffic to the network to obtain traffic attributes upon a link fault. Instead, it leverages readily available metrics (traffic attributes) such as end-to-end delay, packet loss, etc. from the normal traffic generated by users and performs necessary analysis to identify and localize link faults. Traffic attributes are continuously collected and analyzed by a network monitor and analyzer. Thus, the fault manager is able to quickly react upon a change in traffic behavior.

The emergence of machine learning techniques such as deep learning is attracting a great deal of attention in many research and development efforts. Dealing with complex problems is one of the most important advantages of machine learning~\cite{Wang:2018}. With a huge amount of traffic exchanged in the networks, applying machine learning techniques to analyze those traffic attributes could provide useful insights to traffic engineering and fault management. In this paper, we develop a traffic engineering (TE)-based machine learning technique that captures the network state and learns the traffic behavior by passively monitoring the network at egress and ingress nodes. The passive monitoring approach enables a fast link fault identification and localization without additional communication overhead. The traffic features used for machine learning model include the aggregate flow rate, packet loss and round-trip time delay of traffic packets among the monitored nodes. Upon a change in traffic behavior, a link fault could be quickly identified and localized. We consider IoT networks, which could be modeled as complex networks that are able to reflect the randomness and growth characteristics of the real world~\cite{batool:2017,watts1998collective,luo2015small}. We develop a three-stage machine learning-based technique for link fault identification and localization (ML-LFIL). Given traffic information captured from the network, the first stage detects if a link fault has occurred. If a link fault has occurred, the second stage identifies whether a link reconnection has occurred along with a link disconnection or not. Finally, the third stage localizes the link fault, i.e., the location of the disconnected link and/or reconnected link. We train the machine learning model using support vector machine (SVM), random forest (RF) and Multi-Layer Perceptron (MLP, one of the neural network architectures), which have been widely used in the literature for classification and regression problems. We note that training of the learning model can be carried out offline or in a parallel process while fault identification and localization are done by applying the learning model with the data point collected at the time of failure, i.e., a data point is represented by a set of traffic features captured from the network at a time instant. Thus, ML-LFIL enables a fast link fault identification and localization even with large networks. We demonstrate the effectiveness of ML-LFIL in terms of identification and localization accuracy, and fault localization time on different networks, including two random complex networks and a network from the Internet Topology Zoo~\cite{zoo}.     
     
The contributions of the paper are summarized as follows:
  \begin{itemize}
  \item We adopt passive monitoring approach to address the problem of link fault identification and localization in complex networks.
  \item We develop a machine learning model that can learn traffic behavior in normal working condition and in different link fault scenarios.
  \item We develop a TE-based machine learning technique for link fault identification and localization.
  \item We carry out extensive experiments with two random complex networks and one network from the Internet Topology Zoo to demonstrate the effectiveness of the proposed approach.   
 \end{itemize}

The rest of the paper is organized as follows. We review the related
work in Section~\ref{sec:related_work}. We present our proposed
approach in Section~\ref{sec:approach}. We carry out performance
evaluation in Section~\ref{sec:results} before we conclude the paper
in Section~\ref{sec:conclusion}. 

\section{Related Work}
\label{sec:related_work}

\subsection{Network Fault Localization}    

In~\cite{dusia:2016,steinder:2004}, the authors presented a consolidated taxonomy on different techniques that have been developed for localizing link faults. These techniques are broadly categorized as rule-based techniques~\cite{liu:1999,klemettinen:1999}, case-based techniques~\cite{lewis:1993,bennacer:2012}, probability-based techniques~\cite{steinder:2004a,prieto:2011, johnsson:2013, johnsson:2014} and model-based
techniques~\cite{appleby:2002,steinert:2010}. The rule-based techniques rely on a knowledge base developed by the system experts, which is essentially a set of if-then statements i.e., the rules of the system. Nevertheless, these rule-based techniques can neither learn adaptively from the past experience nor from the network dynamics observed from previously unseen traffic behavior. Further, updating and enriching the knowledge base is more complex. Similarly, fault diagnosis by case-based techniques depends on the expert and experience obtained from the past experience. Thus, these solutions can only be applied to those faults that are similar to the ones that have occurred previously. 

In~\cite{johnsson:2013,johnsson:2014}, the authors presented probability-based techniques for fault diagnosis. The location of the link faults in the network is indicated by the corresponding probability mass functions of the links. In comparison with~\cite{steinder:2004a,prieto:2011}, these two works require less computational resources. Model-based techniques~\cite{appleby:2002,steinert:2010} build a mathematical model from a knowledge base to describe the network behaviors. The newly-observed traffic behaviors from the network are compared to those predicted by the model. If the observed behaviors fail to conform to the predicted ones, faults are detected in the network. Thus, the model-based techniques require accurate information about the connections in the network to efficiently diagnose link faults. However, obtaining such information is not always feasible with complex networks owing to their highly-dynamic nature. Differing from all the above techniques, our proposed approach enables a fast fault localization without requiring knowledge about past failures. 
  
\subsection{Machine Learning for Network Fault Management}    

Recently, several works propose to use machine learning techniques for
network fault management~\cite{nguyen:2015,zidi:2018, cheng:2016}. In~\cite{nguyen:2015}, the authors presented a usage-based failure (service disruption) detection method in mobile networks. The authors proposed to monitor aggregated customer usage data and derive a usage pattern for a given geographic region, device
type and service. A drop in aggregated usage (lower than expected)
will be interpreted as a sign of potential service disruption
experienced in that region.  This approach, however, requires the
additional deployment of service monitoring on top of network
monitoring. Further, it requires an accurate user grouping such
that the users in the same group have the same usage pattern. The work presented in~\cite{zhang2018online} uses an online fault detection model using support vector machine (SVM) to detect faults in clouds.
Similarly, the work presented in~\cite{zidi:2018} uses support vector machine (SVM) to classify the received sensor data to detect faults through abnormal behavior in data. This approach requires the traffic to be redirected to the
server where the classifier is deployed, thus incurring a longer delay in
data processing and additional communication overhead. Our approach
requires only the traffic attributes whose size (usually only a few KB)
is much smaller than the data traffic.  

In~\cite{cheng:2016}, the authors proposed a machine learning-based link fault localization coupled with active probing by sending signaling messages to obtain data sets for the machine learning model. Upon a failure, signaling messages are injected into the network with different source and destination node pairs to capture the traffic information such as the number of hops, propagation delay, etc. This information is analyzed by a machine learning model to localize the link fault. On one hand, this incurs additional communication overhead due to signaling messages injected into the network. On the other hand, it incurs an additional delay in fault localization due to the propagation delay of signaling messages across the network. Our work differs from~\cite{cheng:2016} in that we use a passive monitoring approach that identifies and localizes link fault by analyzing the information captured from the normal traffic in normal working conditions and failure scenarios. This enables a fast fault identification and localization without any communication overhead. 

A recent work in the context of 5G networks uses deep learning for link
failure mitigation~\cite{khunteta:2017}. The authors proposed to use
deep learning to analyze the signal conditions of a handover when
mobile devices move from one coverage area to another area under a 
different base station. Based on the signal conditions and the status
of the handovers that happened in the past, the model can classify
whether the handover will be successful or not in advance. Another work
presented in~\cite{ferreira:2017} uses system logs as input for
failure detection and diagnosis for solar-powered wireless mesh
networks. The authors used the knowledge discovery in database
methodology and a pre-defined dictionary of failures based on their
previous experience with the deployment of wireless mesh networks. The
fault detection and diagnosis are solved as a pattern classification
problem. In~\cite{duenas:2018}, the authors described an online
failure prediction system built over Apache Spark that takes a
repository of network management events, trains a random forest model
and uses this model to predict the appearance of future events in near
real time. However, for some failures, e.g., silent failures, no event will
happen in the network, making it hard for the system to detect the failure. Different from these works, we propose to use machine learning techniques to analyze the attributes of normal traffic generated by real users or applications. 

\subsection{Our Work}

Our work addresses the link fault identification and localization problem using passive monitoring. We consider two link fault scenarios: link disconnection and link reconnection. In our previous work~\cite{srinivasan2018te}, we focused on localizing only link disconnections.  In this work, we extend our earlier work by considering both link disconnection and reconnection, which increases the complexity of the problem. Our proposed machine learning approach (ML-LFIL) can achieve high accuracy and  localize link faults faster.

\section{Data Analytics for Link Fault Identification and Localization}
\label{sec:approach}

In this section, we develop our machine learning-based technique for link fault identification and localization (ML-LFIL). In Fig.~\ref{fig:network_architecture}, we present the architecture of ML-LFIL. The network includes both wired and wireless (dashed lines) links. A wireless link can be replaced by a new one with a different end nodes, i.e., a wireless-enabled node can change its existing connection to another node in the network with better wireless channel quality~\cite{kampen2014reconnection}. Due to a link reconnection, certain flows change their paths in the network, thus affecting the end-to-end traffic measurements between the nodes in the network. The network monitor periodically probes the network to capture end-to-end traffic information from flows traversing the network. Traffic information extracted from the flows will be analyzed by the fault manager using a machine learning technique. To avoid the waiting time between instants of traffic sampling, an event-driven approach can be additionally used to trigger link fault identification and localization. When a node experiences abnormal traffic behavior, it sends a request along with traffic measurements for link fault identification and localization. This means that while the monitor keeps probing the network periodically to obtain more data for training the learning model and improving its accuracy, every abnormal traffic behavior will be processed additionally so as to react to link faults in real-time.

\subsection{Traffic Features}    

ML-LFIL carries out link fault identification and localization by analyzing the end-to-end traffic features captured from the network. The accuracy of our machine learning technique depends on the features that are used for training the model. Thus, it is imperative to identify the features to be extracted from network traffic measurements for localizing link disconnections in the network.  At every source/destination node, the following traffic measurements are extracted:

\begin{figure}[t]       
\centering	
\includegraphics[width=0.43\textwidth]{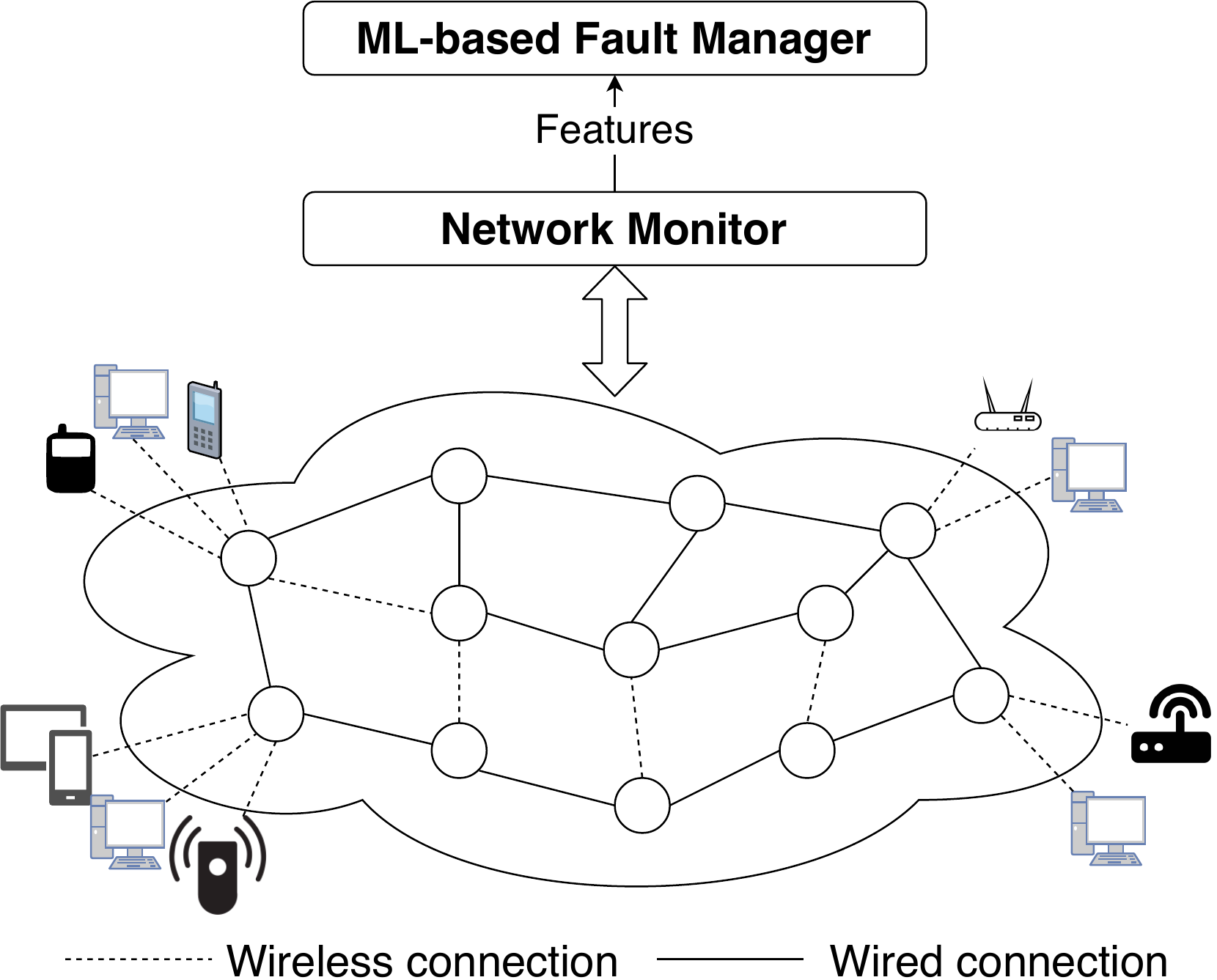}
\caption{Architecture of ML-LFIL.}
\label{fig:network_architecture}  
\vspace{-2ex}
\end{figure}

\begin{itemize}
\item Aggregate transmission rate of flows that destine to other nodes in the network. We denote $b_{s,d}$ as the aggregate rate of the flow that originates from node $s$ to node $d$.
\item End-to-end delay denoted as $d_{s,d}$ that is computed as the round-trip time (RTT) delay of a packet sent from node $s$ to node $d$.
\item Packet loss rate denoted as $l_{s,d}$ is the ratio between the number of packets
  lost on the path between node $s$ and node $d$ and the total number of packets transmitted between two successive sampling instants.
\end{itemize}

While the aggregate rate between every node pair gives us the information about the network load, the end-to-end delay and packet loss features provide indirect information on the path taken by each flow and congestion status in the path. The end-to-end delay can be captured at the hosts in a real network with the timestamp information carried in the packet header. With the same aggregate rate of a flow between a source and destination, longer delay and higher packet loss implicitly mean that a certain link has failed. These traffic measurements captured for different pairs of source and destination nodes help machine learning algorithms to learn the correlation between the learning features to different link fault scenarios. These features are fed to machine learning algorithms as a vector $\mathcal{C}$ as defined below: 
\begin{equation}
     \mathcal{C}=
        \begin{bmatrix}
      	   b_{1,2},b_{1,3},\ldots,b_{i,j},\ldots,b_{V-1, V}, \\
          d_{1,2},d_{1,3},\ldots,d_{i,j},\ldots,d_{V-1,V},\\
           l_{1,2},l_{1,3},\ldots,l_{i,j},\ldots,l_{V-1,V}
        \end{bmatrix} 
       \label{eq:feature-vector1}
\end{equation}
where $V$ is the number of nodes in the network while $b_{s,d}, d_{s,d}$ and $l_{s,d}$ are values of the traffic features defined above. It is to be noted
that the number of features depends on the number of aggregate
flows in the network, which in turn depends on the number of
nodes in the network. Given a network with $V$ nodes, the total number
of aggregate flows denoted as $\mathcal{N}$ is given by $V(V-1)$. We
extract three traffic features for each aggregate flow as
discussed above, and therefore, the size of vector $\mathcal{C}$ is
three times the number of aggregate flows, i.e., $3\mathcal{N} = 3V(V-1)$.  At each sampling instant, a vector $\mathcal{C}$ also called a data point is captured from the network and will be evaluated by ML-LFIL. 

\subsection{ML-LFIL}  
\label{subsec:ML_model}

\begin{figure}[t]       
\centering	
\includegraphics[width =0.48\textwidth]{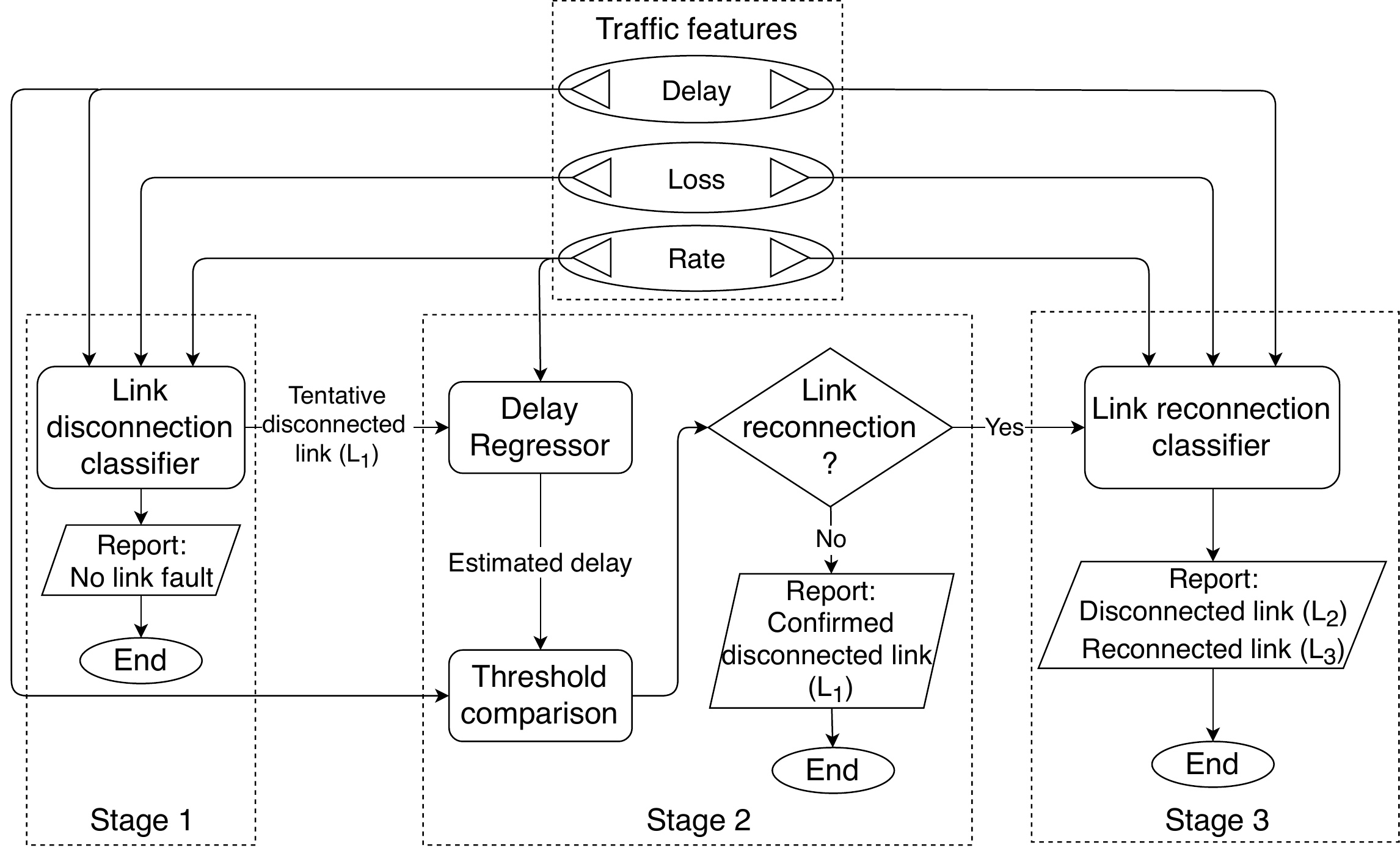}
\caption{Functional block diagram of ML-LFIL.}
\label{fig:machine_learning_model} 
\vspace{-2.7ex} 
\end{figure}

We now present the proposed three-stage machine learning technique for link fault identification and localization (ML-LFIL). In Fig.~\ref{fig:machine_learning_model}, we present the functional block diagram of ML-LFIL. As mentioned earlier, ML-LFIL composes of three stages. Given traffic information captured from the network, the first stage detects if a link disconnection has occurred using a link disconnection classifier. Given that a link disconnection has occurred, the second stage uses a delay regressor to identify the link fault: only a link disconnection has occurred or both link disconnection and link reconnection have occurred. Finally, the third stage uses a link reconnection classifier to localize the link reconnection and may correct the disconnected link resulted by the first stage. Below, we describe the details of each stage.

\subsubsection{Stage 1 -- Link Disconnection Classification}  
\label{subsubsec:class_link_discon}
  
Since different link disconnections may cause different traffic behaviors
represented by the traffic measurements, the problem can be considered as a multi-class machine learning classification problem. The first stage, denoted as ML-LFIL-S1, not only identifies whether a link fault has occurred but also identifies which link has failed. Given a network with $|\mathcal{E}|$ links and a data point, the first stage classifies the data point into one of the link fault classes ($|\mathcal{E}|$ classes) or the ``no-link-fault'' class, where $\mathcal{E}$ is the set of links in the network. Thus, the total the number of classes required for training the machine learning algorithm to detect and localize a link fault in the network is $|\mathcal{E}| + 1$. In our work, we consider a single link fault scenario, i.e., a link fault can be detected and recovered before another fault occurs. This is an acceptable assumption since protection and recovery of the network from multiple simultaneous faults require high-complex algorithms and a large amount of resources to be reserved even when such faults are not frequent. We also note that all the links in the network are equally treated without any priority.

We use all three traffic features and train the learning model using one of the following machine learning algorithms:
\begin{itemize}

\item Support Vector Machine (SVM)~\cite{suykens1999least} is a supervised machine learning technique that tries to separate data points into two different classes by identifying the best possible separating hyperplane. It can be extended to multi-class problems by constructing multiple hyperplanes.

\item Multi-Layer Perceptron (MLP)~\cite{gardner1998artificial} is a class of artificial neural networks. The number of layers and the number of neurons in each layer depends on the complexity of the machine learning problem. A back-propagation algorithm is used by MLP for training and obtaining the weights of the neurons in the neural network.  
  
\item Random Forest (RF)~\cite{kulkarni2013random} is a classifier algorithm that constructs multiple decision trees during the training phase and outputs the mode of the individual trees as the class label. It suits well for multi-class classification problems as the link fault identification and localization.
\end{itemize}

We note that while there exist many machine learning algorithms that can be used for a classification problem, SVM, RF and MLP have demonstrated their best performance over other algorithms~\cite{srinivasan2018te}. Hence, in this work, we only adopt these three machine learning algorithms to train our learning model to achieve the best performance.

The output of the first stage is a {\it tentative} disconnected link ($L_1$) in case a link disconnection has occurred or a message stating the normal working condition of the network if no link fault has occurred. The ``tentative'' term means that the disconnected link $L_1$ may not be accurately determined due to a link reconnection that has occurred along with the link disconnection but causing similar behavior as in case of link disconnection alone. Even though $L_1$ may be wrongly classified, it shows ML-LFIL is sensitive with link fault and it triggers the other two stages for further analysis to identify and correctly localize the link fault. 

\subsubsection{Stage 2 -- Link Fault Identification (ML-LFIL-S2)}
\label{subsubsec:reg_link_fail}

To identify the link fault, we estimate the end-to-end delay of the network traffic caused by the disconnection of the tentative link $L_1$, using aggregate flow rates captured from the network. The estimated end-to-end delay is compared with the actual delay captured from the network. We use the mean square error to compute the difference between the estimated delay and actual delay. If the difference is less than the threshold value (say $10\%$), we can confirm that link reconnection has not occurred along with the link disconnection and $L_1$ is the exact link that has been disconnected. The rationale behind is that if only a link disconnection has occurred, the flows affected by the disconnected link have to traverse a long path, thus experiencing a longer delay, compared to the case where the disconnected link is replaced by a new link. 

We develop a regression learning model to estimate the end-to-end delay of all the flows in the network. The model is trained with the aggregate flow rates and the tentative disconnected link $L_1$. These features are fed to machine learning algorithms as vector $\mathcal{R}$ given by, 
\begin{equation}
     \mathcal{R}=  
        \begin{bmatrix}
      	   b_{1,2},b_{1,3},\ldots,b_{i,j},\ldots,b_{V-1, V},s_l,d_l \\
        \end{bmatrix} 
       \label{eq:feature-vector2}
\end{equation}
where $V$ is the number of nodes in the network, $b_{i,j}$ is aggregate flow rate between node $i$ and node $j$, and ($s_l, d_l$) are source and sink of the tentative disconnected link $L_1$.  We train the regression model using MLP. In this work, we use an MLP with $3$ hidden layers and $400$ neurons in each hidden layer with ReLU activation functions. We note that packet loss information can also be used as an input feature to estimate the end-to-end delay. However, this unnecessarily increases the complexity of the problem and design of MLP.

\subsubsection{Stage 3 -- Link Reconnection Classification}  
\label{subsubsec:class_link_rec}    

Given that a link reconnection has been identified by the second stage, the third stage of ML-LFIL localizes both disconnected link  ($L_2$) and reconnected link  ($L_3$) using a link reconnection classifier. The disconnected link $L_2$ might be different or the same as the tentative disconnected link $L_1$ depending on the accuracy of the link disconnection classifier. Similar to link disconnection classification, link reconnection classification in the third stage of ML-LFIL (ML-LFIL-S3) is also a multi-class machine learning classification problem. However, each class in this problem includes a pair of a disconnected link and a reconnected link. All the three traffic features are used in the learning model that is trained using SVM, MLP or RF.    

\subsection{Illustrative Example} 

\begin{figure}[t]      
\centering	
\begin{subfigure}[b]{0.235\textwidth}  
\centering
\includegraphics[width=\textwidth]{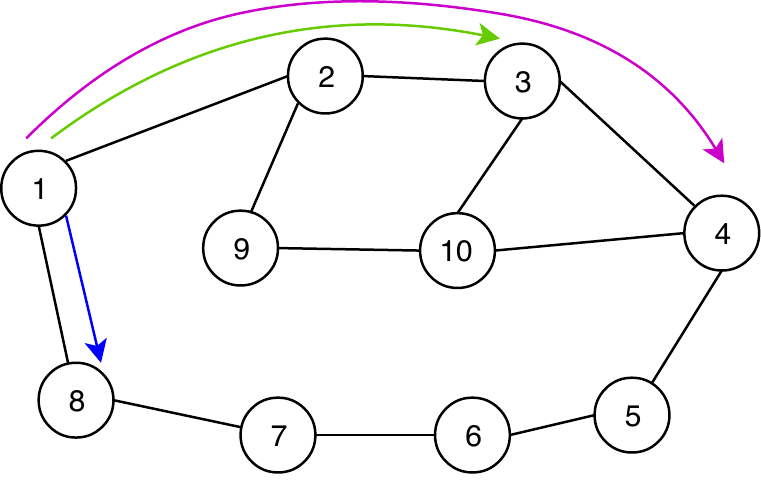}
\caption{Normal scenario.}
\label{fig:Normal_illustration}
\vspace{0.5ex}  
\end{subfigure}      
\hspace{4px} 
\begin{subfigure}[b]{0.235\textwidth}
\centering
\includegraphics[width=\textwidth]{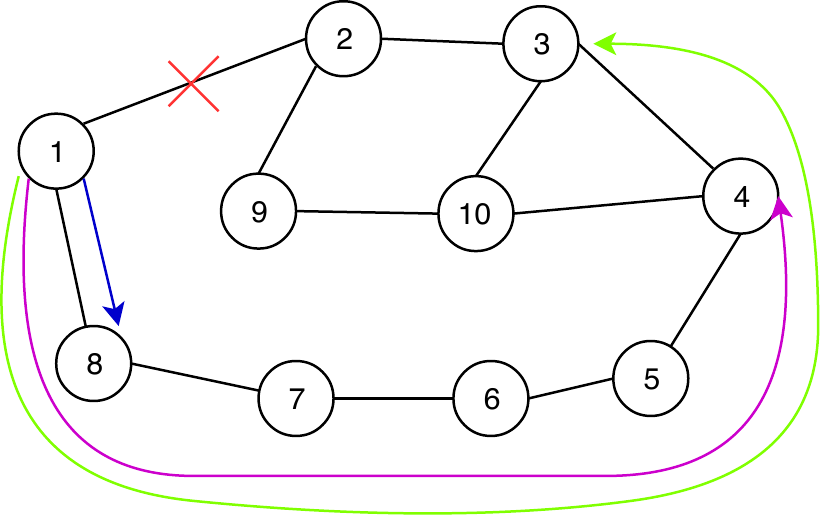}
\caption{($1-2$) disconnects.}
\label{fig:l1fl_illustration}
\vspace{-0.2ex}  
\end{subfigure}  
\begin{subfigure}[b]{0.235\textwidth}      
\centering
\includegraphics[width=\textwidth]{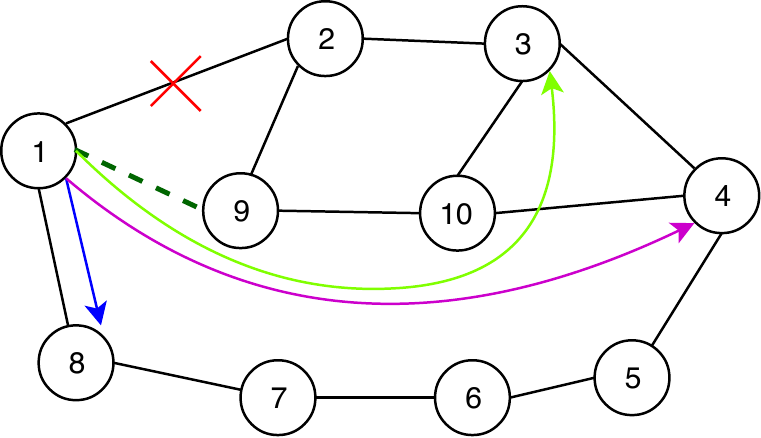}
\caption{($1-2$) is replaced by ($1-9$).}
\label{fig:l1rc_illustration}
\end{subfigure}  
\vspace{-0.2ex}  
\caption{Illustrative example.}   
\label{fig:illustration}  
\vspace{-3ex}
\end{figure}

We illustrate the working of ML-LFIL with an example. Given a $10$-node network as depicted in Fig.~\ref{fig:illustration}, all the flows are routed on the shortest paths. Consider three flows with source and destination as $\langle{}1, 3\rangle$, $\langle{}1, 4\rangle$ and $\langle{}1, 8\rangle$. Fig.~\ref{fig:Normal_illustration} depicts the paths traversed by the flows in normal working conditions. Upon disconnection of the link ($1-2$), the affected flows are rerouted and the paths traversed by the affected flows are depicted in Fig.~\ref{fig:l1fl_illustration}. We can see that the two affected flows $\langle{}1, 3\rangle$ and $\langle{}1, 4\rangle$ are rerouted through alternate paths, which are longer than the paths used in normal working conditions, leading to longer propagation delay. Further, all of the flows now traverse through the same link ($1-8$) and share a limited amount of bandwidth. Thus, they experience additional delay due to congestion on link ($1-8$), which is a sign of link fault. Similarly, packet loss is a useful measurement to identify link disconnections. Upon a link disconnection, all the packets sent through the disconnected link are dropped before an alternate path is found, thus leading to increased packet loss. The longer the time needed for the network to find an alternate path, the higher the packet loss. In the scenario shown in Fig.~\ref{fig:l1fl_illustration}, upon disconnection of link ($1-2$), the two flows  $\langle{}1, 3\rangle$ and $\langle{}1, 4\rangle$ experience higher packet loss. 

Fig.~\ref{fig:l1rc_illustration} depicts routing solutions when link ($1-2$) is replaced by link ($1-9$). Except flow $\langle{}1, 8\rangle$,  both flows $\langle{}1, 3\rangle$ and $\langle{}1, 4\rangle$ traverse through different paths. This change in the routing path of the flows affects traffic measurements of the flows. However, the effect might not be significant when compared to the case of link disconnection alone. Indeed, even though link ($1-2$) is disconnected, the end-to-end delay of flow $\langle{}1,4\rangle$ on the alternate path remains unchanged ($3$ hops) and the alternate path of flow $\langle1,3\rangle$ increases only $1$ hop. Thus, the link disconnection classifier (the first stage of ML-LFIL) might not be able to correctly localize the disconnected link. Using the regression model to estimate the end-to-end delay of flows given a link fault allows us to infer whether a link reconnection has occurred, thus being able to localize both disconnected link and reconnected link.

\section{Performance Study}
\label{sec:results}

\subsection{Simulation Settings and Data Collection}
\label{subsec:simulation_settings}

We implement ML-LFIL and carry out experiments to evaluate its performance using the Mininet platform. We consider two complex network topologies: a $30$-node network with $36$ links and a $60$-node network with $68$ links. The two networks are created based on the small-world complex network model that emulates each network node with $4$ neighbors and the probability of adding another edge for each node is $0.35$. The traffic between the nodes in the networks is generated using \textit{iperf3} tool with the rates between each node pair randomly chosen in the range $[1,300]$ Mbps. We also evaluate our proposed ML-LFIL method with the Interroute topology with $100$ nodes and  $120$ links with a realistic traffic trace. We use Wireshark to capture the traffic measurements, i.e., aggregate flow rate, end-to-end delay and packet loss in normal working conditions and different link fault scenarios. To emulate link disconnections, we randomly remove a link in the network topology in Mininet platform while traffic flows are being forwarded across the network. All the affected flows will be rerouted through alternate paths. We collect the traffic measurements for multiple link disconnection scenarios to train the machine learning model in the first two stages of ML-LFIL (link disconnection classification and link fault identification). Similarly, to emulate link reconnections, we randomly remove one link and add to the network a new link that has same source node as the removed link. Different link reconnection scenarios are generated by removing and adding different links between different nodes in the network. The traffic measurements captured with different link reconnection scenarios are used to train the third stage of ML-LFIL (link reconnection classification). Following the data collection, the data is preprocessed using normalization techniques and Principal Component Analysis (PCA) to enable better performance of machine learning algorithms.  

For the $30$-node topology, we train the link disconnection classifier in the first stage and the delay regressor in the second stage of ML-LFIL with $28,000$ data points for each class of link disconnections, i.e., we use about $800,000$ data points in the training data set. The test data set includes $200,000$ data points. The link reconnection classifier in the third stage of ML-LFIL is trained and tested using $500,000$ data points and $100,000$ data points, respectively. For the $60$-node topology, we train the link disconnection classifier and the delay estimator with $60,000$ data points for each class, i.e., about $3,200,000$ data points in the training data set. The test data set has $800,000$ data points. Similarly, we train and test the link reconnection classifier with $800,000$ and $200,000$ data points, respectively. For the 100-node Interroute topology, we train the link disconnection classifier (ML-LFIL-S1) and the delay estimator (ML-LFIL-S2) with $8,000$ data points for each class, i.e., about $672,000$ data points in the training data set. The test data set has $160,000$ data points. We train and test the link reconnection classifier with $600,000$ and $100,000$ data points, respectively.

\subsection{Performance Metrics}

We use the following performance metrics to evaluate the performance
of different machine learning algorithms: 
\begin{itemize}

\item Precision: The ratio of the number of link faults correctly classified over the total number of data points classified as faults. The precision value is computed as follows: 
\begin{equation}
	\mathcal{P} = \dfrac{T_P}{T_P +F_P} 
\label{eq:precision}
\end{equation}
where $\mathcal{P}$ is the precision value, $T_P$ is the number of
``true positives''	and $F_P$ is the number of ``false
positives''. 

\item Recall:  The ratio of the number of data points associated with link
  faults correctly classified over the total number of
  data points associated with link faults that have occurred. The
  recall value is given by: 
\begin{equation}
	\mathcal{R} = \dfrac{T_P}{T_P +F_N} 
\label{eq:recall}
\end{equation}
where $F_N$ is the number of ``false negatives''.	

\item $F_1$-Score: The $F_1$-Score is the harmonic average of the
  precision and recall values. It takes a value in the range $[0,1]$. The higher the value of $F_1$-Score, the better the performance of the machine learning technique, i.e., we obtain perfect precision and recall values when $F_1$-Score reaches $1$. It is computed as
  follows:  
\begin{equation}
	F_{1}\text{-Score} = \dfrac{2\mathcal{P}\mathcal{R}}{\mathcal{P} +\mathcal{R}}. 
\label{eq:f1-score}
\end{equation}

\item $R^2$-Score: The goodness of the MLP used in the delay regressor (ML-LFIL-S2). It takes the values in the range $[0,1]$. The higher the value of $R^2$ score, the better the performance of the learning model.	
\item Fault detection accuracy: The ratio of the number of data points
  associated with link faults detected (regardless of the correctness of the tentative disconnected link) over the total number of data points associated with link faults that have actually occurred.

\item Fault localization time: The time is taken to localize the link upon its fault. We compare ML-LFIL with a ping-based active probing approach that sends signaling messages to all the nodes in the network to obtain traffic information before analyzing to localize the link fault.
\end{itemize}

\subsection{Analysis of Results}
\label{subsec:results_analysis}

\subsubsection{Performance with the $30$-node network}
\label{subsec:30 node network}
        
In this section, we evaluate the performance of ML-LFIL with the $30$-node network.  We first evaluate the performance of the first stage of ML-LFIL, denoted as ML-LFIL-S1since its performance affects the overall performance of ML-LFIL. Indeed, if it cannot classify the traffic features associated with a link fault and returns a ``no-link-fault'' message, ML-LFIL will stop without further analysis. We first consider the fault scenarios where only link disconnections are present. Both the precision and recall values on the training data set are close to $100\%$ for all the machine learning algorithms (SVM, MLP and RF) used to train ML-LFIL-S1. This demonstrates that the link disconnection classifier in ML-LFIL-S1 has been well trained.

\begin{figure}[t]  
  \centering
  \includegraphics[width = 0.43\textwidth]{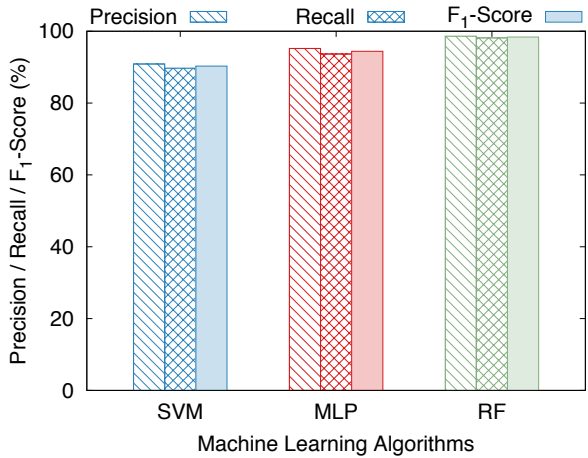}
  \caption{Precision/Recall/$F_1$-Score  of ML-LFIL-S1 in $30$-node network with only link disconnections.}
  \label{fig:prec_rec_30}
    \vspace{-1.5ex} 
\end{figure}

In Fig.~\ref{fig:prec_rec_30}, we plot the precision, recall and $F_1$-Score values of all the algorithms used to train ML-LFIL-S1 on the test data set. The results show that ML-LFIL-S1 achieves high performance of at least $90\%$ of precision, recall and $F_1$-Score values. The results also show that all the algorithms have a high precision value. This means that all the algorithms have a low false positive rate. Similarly, the high recall value implies a low false negative rate. We can observe from Fig.~\ref{fig:prec_rec_30} that there is a lower false positive rate compared to the false negative rate.  Among the machine learning algorithms, RF algorithm outperforms SVM and MLP algorithms with a precision of about $98.6\%$, a recall of about $98.1\%$ and $F_1$-Score of about $98.4\%$. This shows that the learning model trained with RF algorithm classifies link disconnection with minimal misclassification or noise.  

\begin{figure}[t]
  \centering
  \includegraphics[width = 0.43\textwidth]{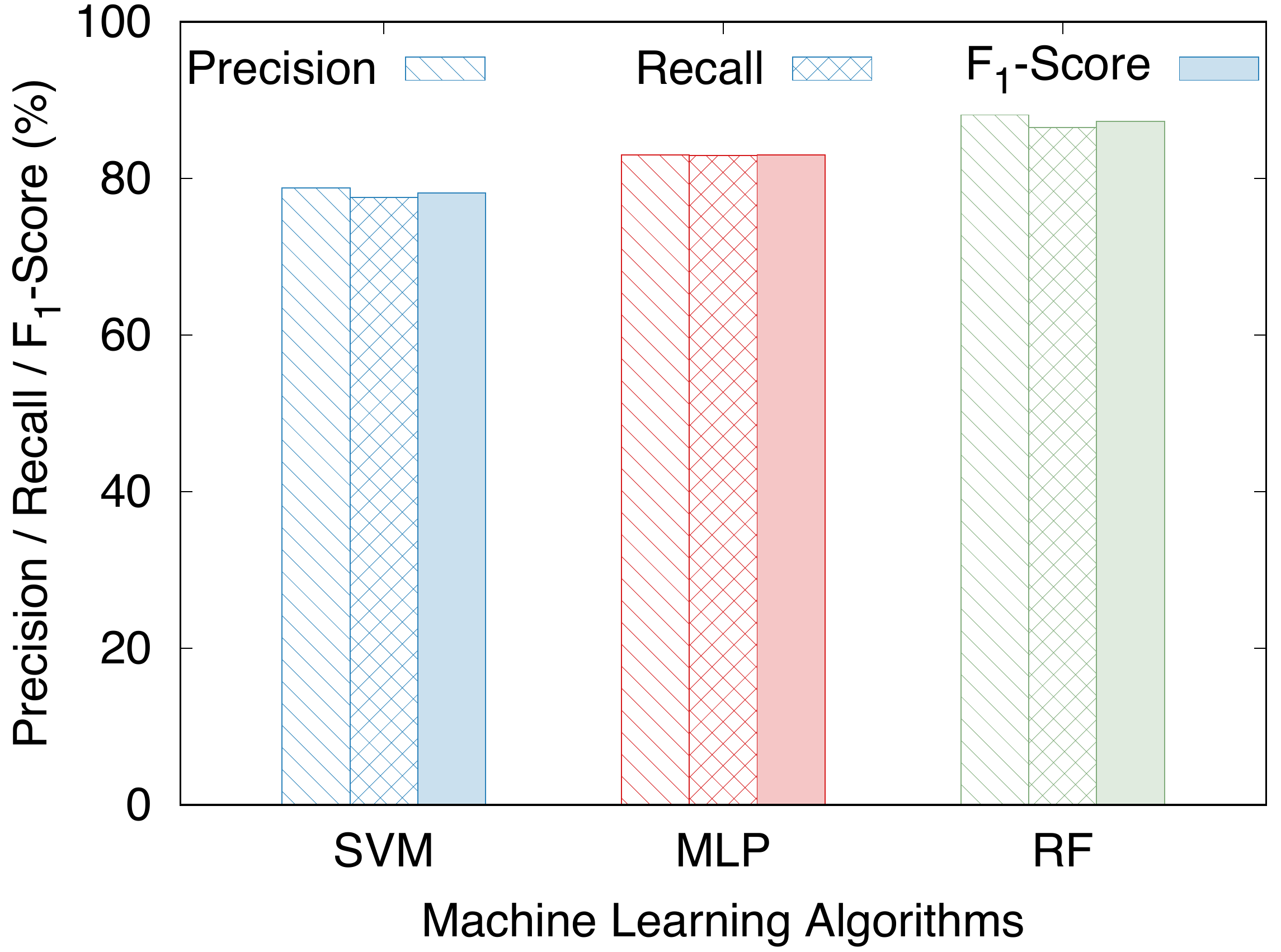}
  \caption{Precision/Recall/$F_1$-Score of ML-LFIL-S1 in $30$-node network with both link disconnections and link reconnections.} 
  \label{fig:prec_rec_30lr}
    \vspace{-2.5ex} 
\end{figure}	

\begin{figure}[t]  
  \centering
  \includegraphics[width = 0.43\textwidth]{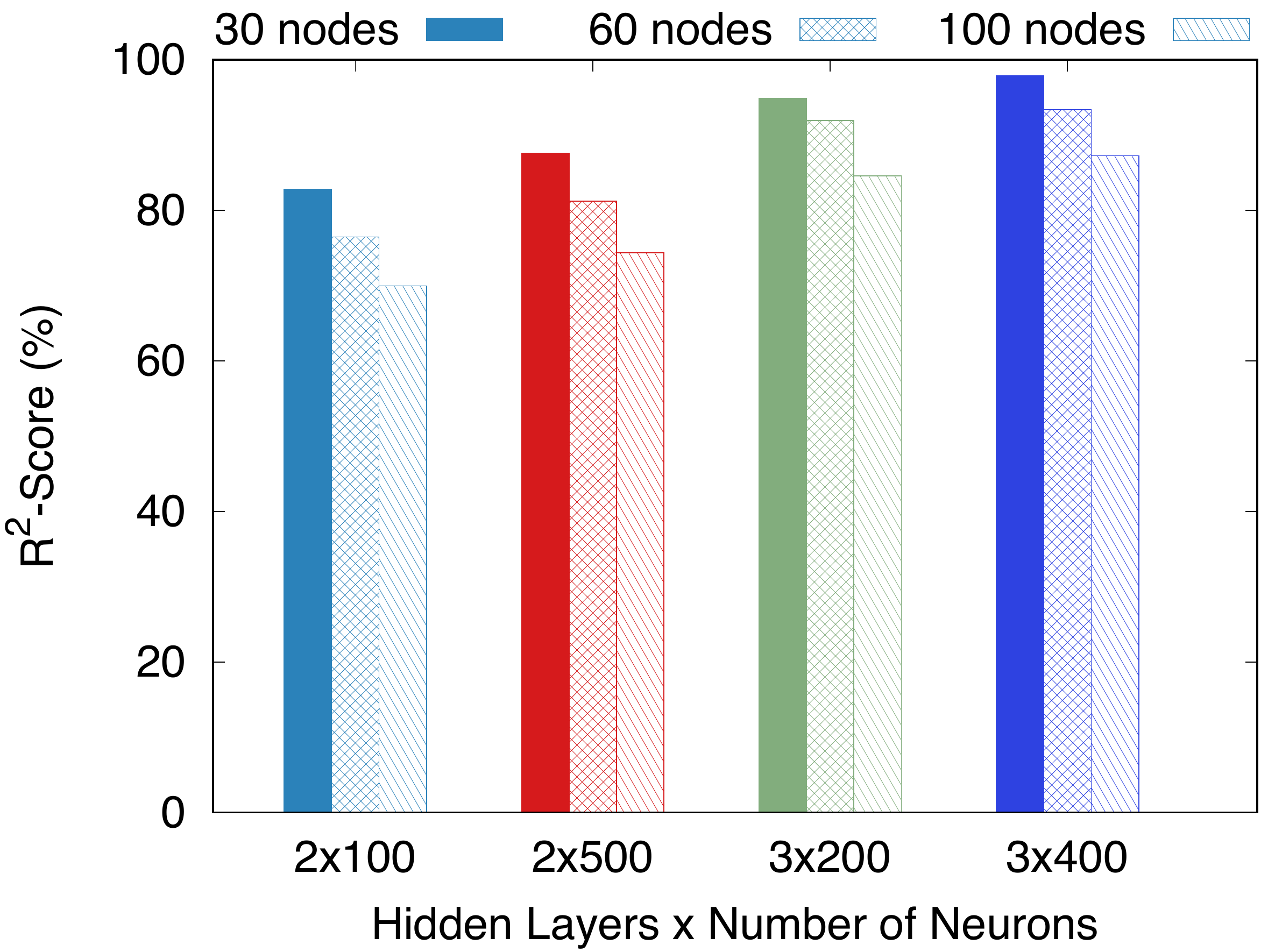}
  \caption{Delay regressor accuracy  of MLP in ML-LFIL-S2 in $30$-node network with only link disconnections for different hidden layers and hidden neurons.}
  \label{fig:MLP_30}
    \vspace{-1ex} 
\end{figure}	

\begin{figure}[t]  
  \centering
  \includegraphics[width = 0.43\textwidth]{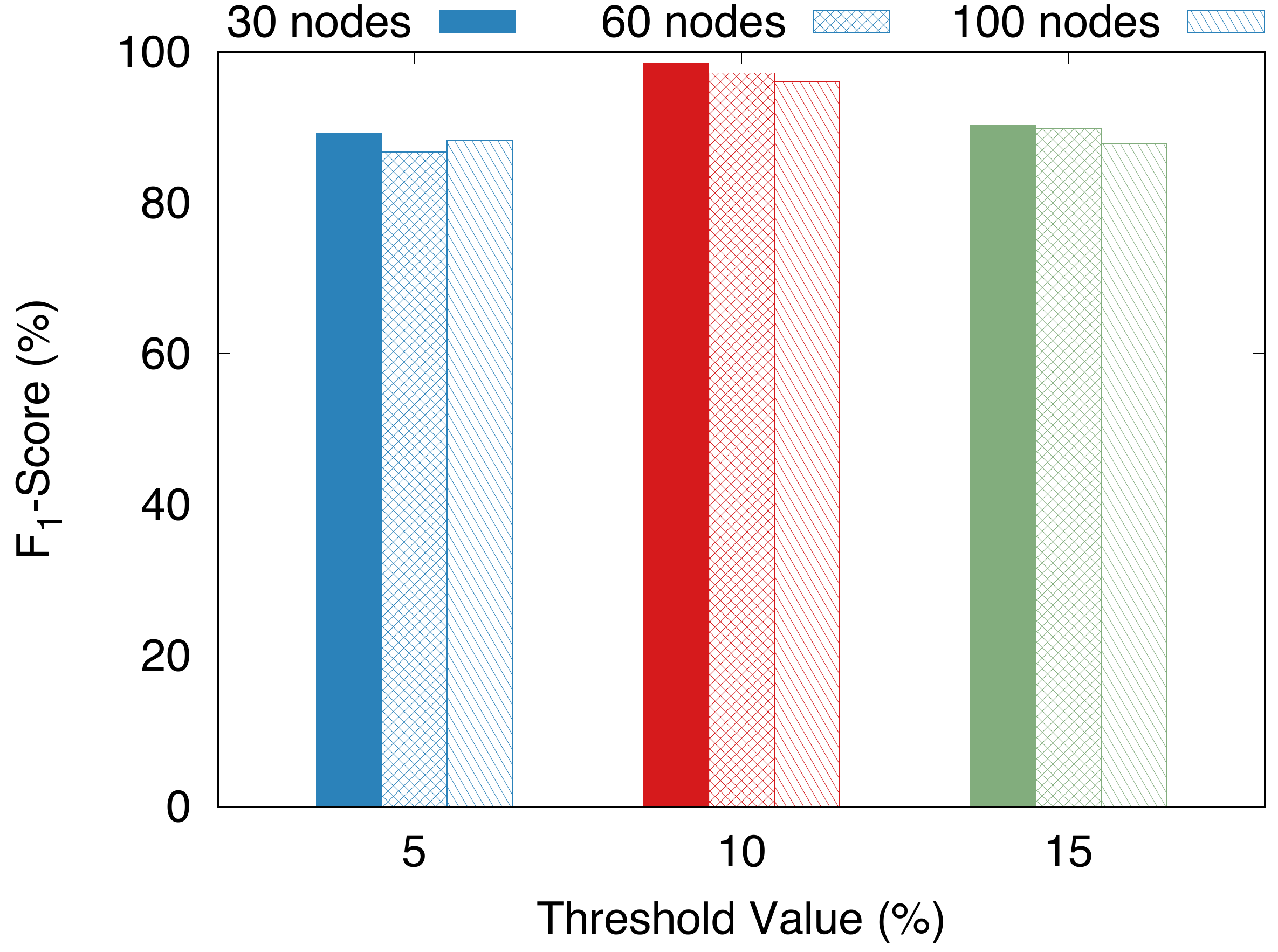}
  \caption{Link disconnection/reconnection accuracy  of ML-LFIL-S2 in $30$-node network for different threshold values.}
  \label{fig:threshold_30}
   \vspace{-2ex} 
\end{figure}	

\begin{figure}[t]
  \centering
  \includegraphics[width = 0.43\textwidth]{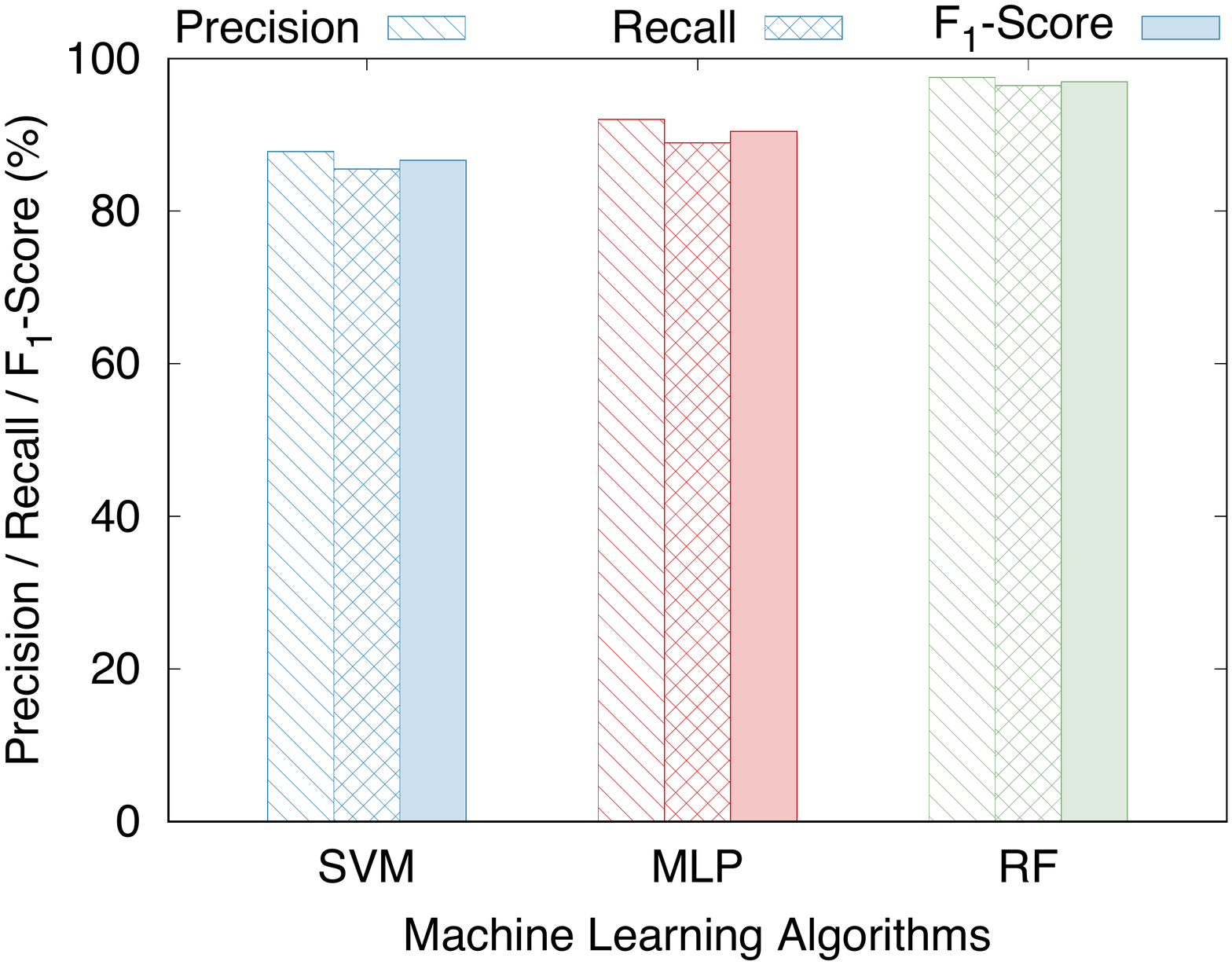}
  \caption{Precision/Recall/$F_1$-Score of ML-LFIL in $30$-node network with both link disconnections and reconnections.}
  \label{fig:prec_rec_30lrcr}
    \vspace{-1.5ex}    
\end{figure}

We now consider the fault scenarios where both link disconnections and link reconnections are present. We observe that the precision, recall and $F_1$-Score values of all the algorithms decrease due to misclassification as shown in Fig.~\ref{fig:prec_rec_30lr}. This is because the link disconnection classifier in ML-LFIL-S1 is trained only with the traffic features captured from the link disconnection scenarios. Thus, it will misclassify a data point if the traffic behavior in a link reconnection scenario is similar to that in a link disconnection scenario. Nevertheless, RF algorithm always outperforms other algorithms with a precision of $88.1\%$, a recall of $86.5\%$ and an $F_1$-Score of $87.3\%$. This explains why we develop the regression model in the second stage of ML-LFIL to identify the link fault whether or not a link reconnection has occurred. We note that since RF has the best performance among the machine algorithms used to train ML-LFIL-S1, when evaluating the performance of the subsequent stages, we use RF for the first stage, ML-LFIL-S1.

In Fig.~\ref{fig:MLP_30}, we present the accuracy for different MLP architectures, i.e., different number of hidden layers and number of units in each hidden layer. MLP with 3 hidden layers and 400 neurons perform better than the other combinations, having an $R^2$-Score of $98\%$, $93\%$ and $87\%$ for the 30-node, 60-node and 100-node Interroute topologies, respectively. We can have even a deeper and larger neural network, but it would lead to similar accuracy and much more complexity and thus, would be a overkill for our problem. Using the best architecture of MLP with with 3 hidden layers and 400 neurons, in Fig.~\ref{fig:threshold_30}, we present the accuracy of ML-LFIL-S2 in identifying the link fault, i.e., whether a link reconnection has occurred or not, for different threshold values. It can be seen that the $F_1$-Score is maximum for the threshold value of $10\%$ in the threshold comparator module, with $98.5\%$, $97.2\%$ and $96\%$ for the 30-node, 60-node and 100-node Interroute topologies, respectively. Having a high threshold value will lead to link reconnections go undetected and to be identified as just link disconnections, whereas having a low threshold value will lead to link disconnections being identified as link reconnections. Thus, the threshold value is crucial in distinguishing between link reconnections and link disconnections in the network. We use the threshold value of $10\%$ for the remaining experiments.

When we use ML-LFIL with all three stages on the test data set that includes both link disconnections and link reconnections, we obtain a significant performance improvement. In Fig.~\ref{fig:prec_rec_30lrcr}, we present the precision, recall and $F_1$-Score of ML-LFIL. We note that, since we need to have high accuracy in the initial stages to achieve high accuracy of the overall model, we present the results of the overall ML-LFIL model, by evaluating the third stage with different classifiers in Fig.~\ref{fig:prec_rec_30lrcr}, with the first stage trained with RF and the second stage trained with MLP. The results show that RF algorithm always outperforms the other algorithms with $97.5\%$ of precision, $96.46\%$ of recall and $97\%$ of $F_1$-Score. It is to be noted that we use the same algorithm for both link disconnection classifier in the first stage and link reconnection classifier in the third stage of ML-LFIL. SVM algorithm has the worst performance among the algorithms. Nevertheless, it achieves $87.8\%$ of precision, $85.5\%$ of recall and $86.7\%$ of $F_1$-Score. 

\subsubsection{Performance with the $60$-node network}  
\label{subsec:60 node network} 

\begin{figure}[t]
  \centering
  \includegraphics[width = 0.43\textwidth]{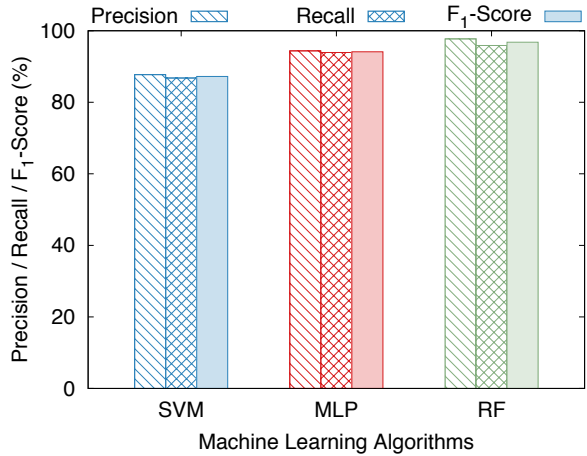}
  \caption{Precision/Recall/$F_1$-Score  of ML-LFIL-S1 in $60$-node network with only link disconnections.}
  \label{fig:prec_rec_60}
 \vspace{-2ex}
\end{figure}	

In this section, we evaluate the performance of ML-LFIL with a $60$-node network. Similar to the analysis with the $30$-node topology, we evaluate the performance of  ML-LFIL-S1 and then ML-LFIL as a whole. In Fig.~\ref{fig:prec_rec_60}, we present the precision, recall and $F_1$-Score values of ML-LFIL-S1 for the test data set that includes only link disconnections. We obtain similar performance trend as shown in the previous section. It is evident that RF algorithm performs the best with $97.7\%$, $95.8\%$ and $96.7\%$ of precision, recall and $F_1$-Score, respectively.

\begin{figure}[t]
  \centering
  \includegraphics[width = 0.43\textwidth]{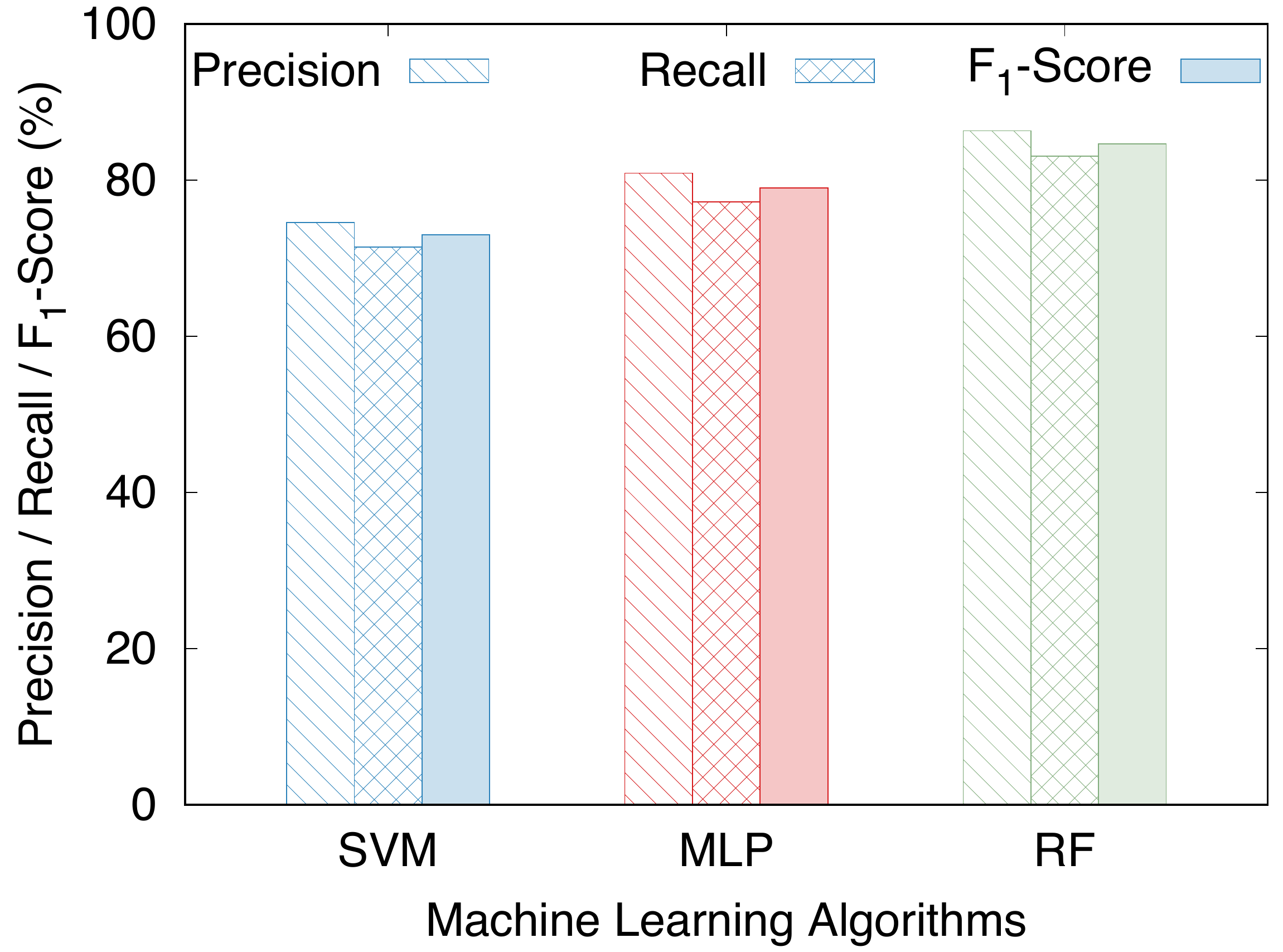}
  \caption{Precision/Recall/$F_1$-Score of ML-LFIL-S1 in $60$-node network with both link disconnections and link reconnections.}
  \label{fig:prec_rec_60lr}
  \vspace{-1.5ex}
\end{figure}	

\begin{figure}[t]  
  \centering
  \includegraphics[width = 0.43\textwidth]{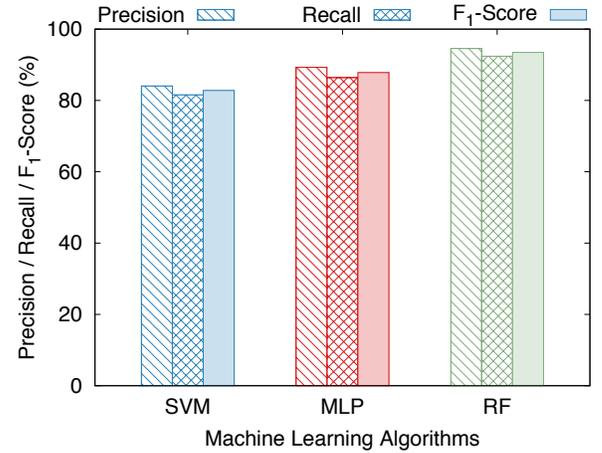}
  \caption{Precision/Recall/$F_1$-Score of ML-LFIL in $60$-node network with both link disconnections and reconnections.}
  \label{fig:prec_rec_60lrcr}
  \vspace{-1ex}
\end{figure}	

When considering both link disconnections and link reconnections, we also observe a performance degradation due to the misclassification. As shown in Fig.~\ref{fig:prec_rec_60lr}, RF attains only $86.3\%$, 	$83.06\%$ and $84.65\%$ for precision, recall and $F_1$-Score, respectively.  In Fig.~\ref{fig:prec_rec_60lrcr}, we present the precision, recall and $F_1$-Score of all the three algorithms used to train the third stage of ML-LFIL in the presence of both link disconnections and link reconnections. We obtain high precision and recall values of $94.56\%$ and $92.4\%$ for ML-LFIL with the RF algorithm. Similarly, we  obtain high $F_1$-Score of at least $82.8\%$ for SVM and $93.47\%$ for RF algorithm. This demonstrates the effectiveness of ML-LFIL in identification and localization of link faults even with large-scale complex networks.  

\subsubsection{Performance with the $100$-node Interroute network}
\label{subsec:100 node network} 
\begin{figure}[t]
  \centering
  \includegraphics[width = 0.43\textwidth]{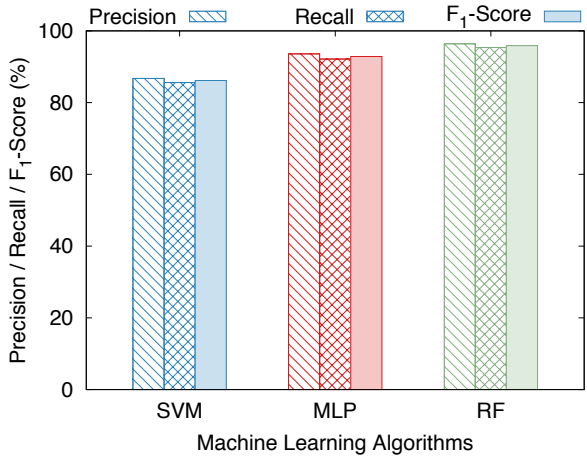}
  \caption{Precision/Recall/$F_1$-Score  of ML-LFIL-S1 in the Interroute network with only link disconnections.}
  \label{fig:prec_rec_100}
 \vspace{-1.5ex}
\end{figure}	

In this section, we evaluate the performance of ML-LFIL with the Interroute network with $100$ nodes. Similar to the analysis with the previous two topologies, we evaluate the performance of  ML-LFIL-S1 and then ML-LFIL as a whole. In Fig.~\ref{fig:prec_rec_100}, we present the precision, recall and $F_1$-Score values of ML-LFIL-S1 for the test data set that includes only link disconnections. We obtain similar performance trend as seen in the previous sections. It is evident that RF algorithm performs the best with $96.37\%$, $95.37\%$ and $95.87\%$ of precision, recall and $F_1$-Score, respectively.

\begin{figure}[t]
  \centering
  \includegraphics[width = 0.43\textwidth]{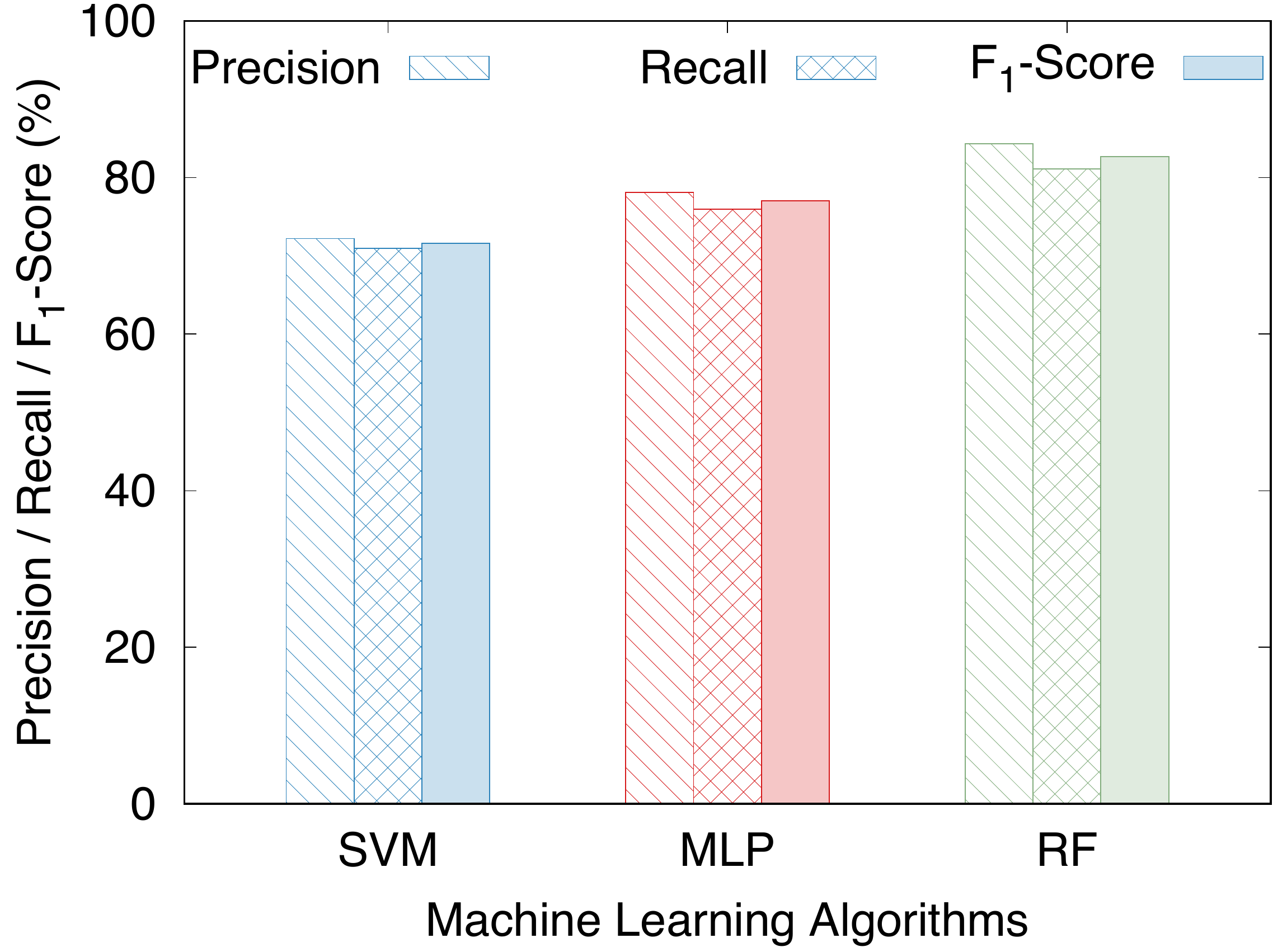}
  \caption{Precision/Recall/$F_1$-Score of ML-LFIL-S1 in the Interroute network with link disconnections and reconnections.}
  \label{fig:prec_rec_100lr}
  \vspace{-2ex}
\end{figure}	

\begin{figure}[t]  
  \centering
  \includegraphics[width = 0.43\textwidth]{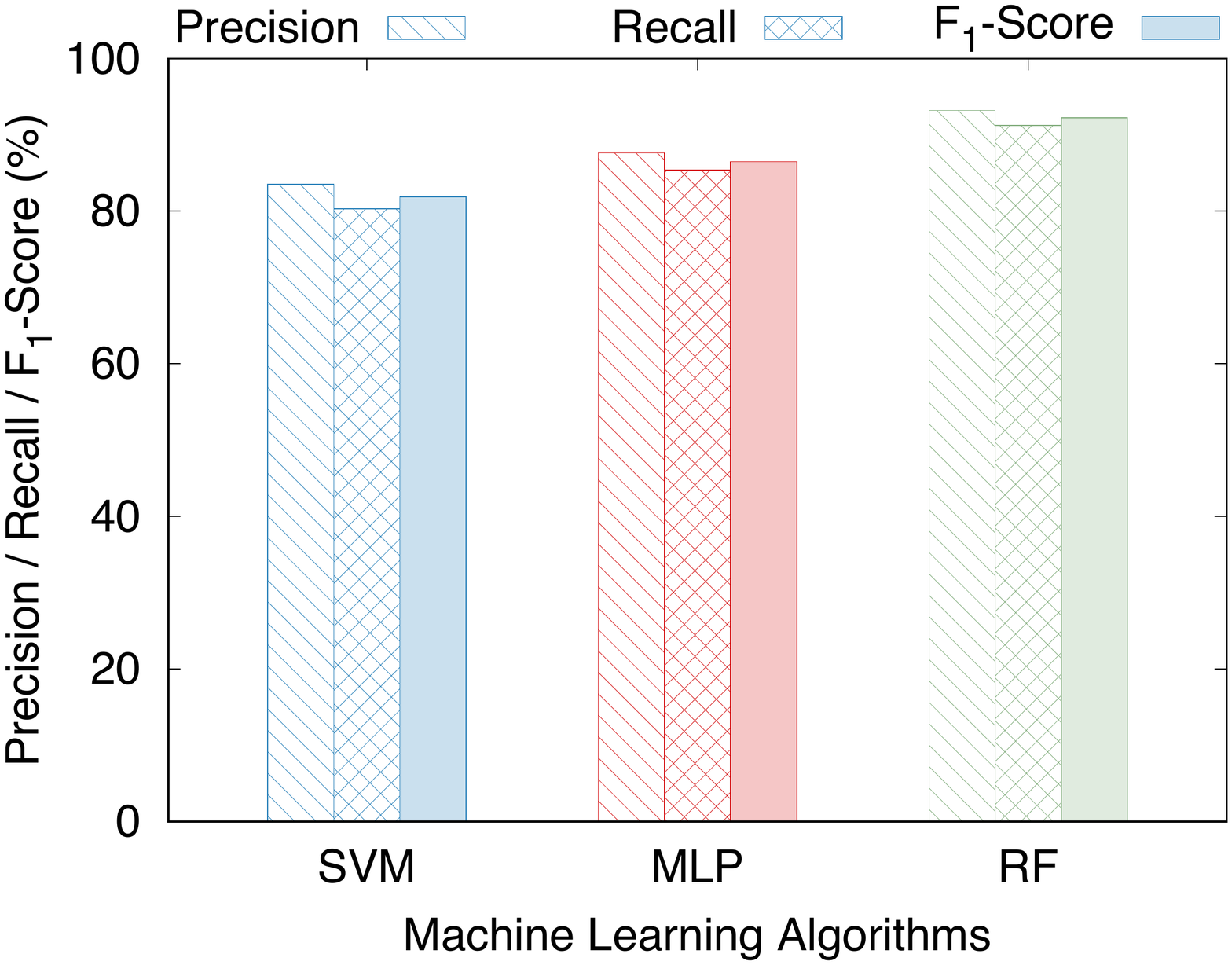}
  \caption{Precision/Recall/$F_1$-Score of ML-LFIL in the Interroute network with link disconnections and reconnections.}
  \label{fig:prec_rec_100lrcr}
  \vspace{-1ex}
\end{figure}	

\begin{table}[t]
\centering
\caption{Summary of ML-LFIL performance with the test dataset for the three different topologies (in \%)}
\begin{tabular}{ | l | c | c | c |} 
\hline
\textbf {Topology} & \textbf {Precision} & \textbf{Recall } & \textbf{$F_1$-Score} \\ 
  \hline
  \hline
$30$ nodes & 97.52  & 96.46 & 97.00 \\
\hline
$60$ nodes & 94.56 & 92.40 & 93.47\\
\hline
$100$ nodes& 93.20 & 91.22 & 92.20\\
\hline
\end{tabular}
\label{table:summary_comp}
\vspace{-2ex}
\end{table} 

When considering both link disconnections and link reconnections, we also observe a performance degradation due to the misclassification. As shown in Fig.~\ref{fig:prec_rec_100lr}, RF attains only $84.3\%$,  $81.1\%$ and $82.7\%$ for precision, recall and $F_1$-Score, respectively.  With the output from ML-LFIL-S1 trained by RF and ML-LFIL-S2 trained by MLP, in Fig.~\ref{fig:prec_rec_100lrcr}, we present the precision, recall and $F_1$-Score of all the three algorithms used to train the third stage of ML-LFIL in the presence of both link disconnections and link reconnections. We obtain high precision and recall values of $93.2\%$ and $91.22\%$ for ML-LFIL with the RF algorithm. We also obtain high $F_1$-Score of at least $81.87\%$ for SVM and $92.2\%$ for RF algorithm. This demonstrates the effectiveness of ML-LFIL in identification and localization of link faults even with large networks and realistic traffic traces. A summary of precision, recall and $F_1$-Score values with the test data sets for all the three networks, using the best algorithm at each stage of ML-LFIL is given in Table~\ref{table:summary_comp}.

\subsubsection{Fault Detection Accuracy}
\label{subsec:Fault Detection Accuracy}
\begin{figure}[t]
  \centering
  \includegraphics[width = 0.43\textwidth]{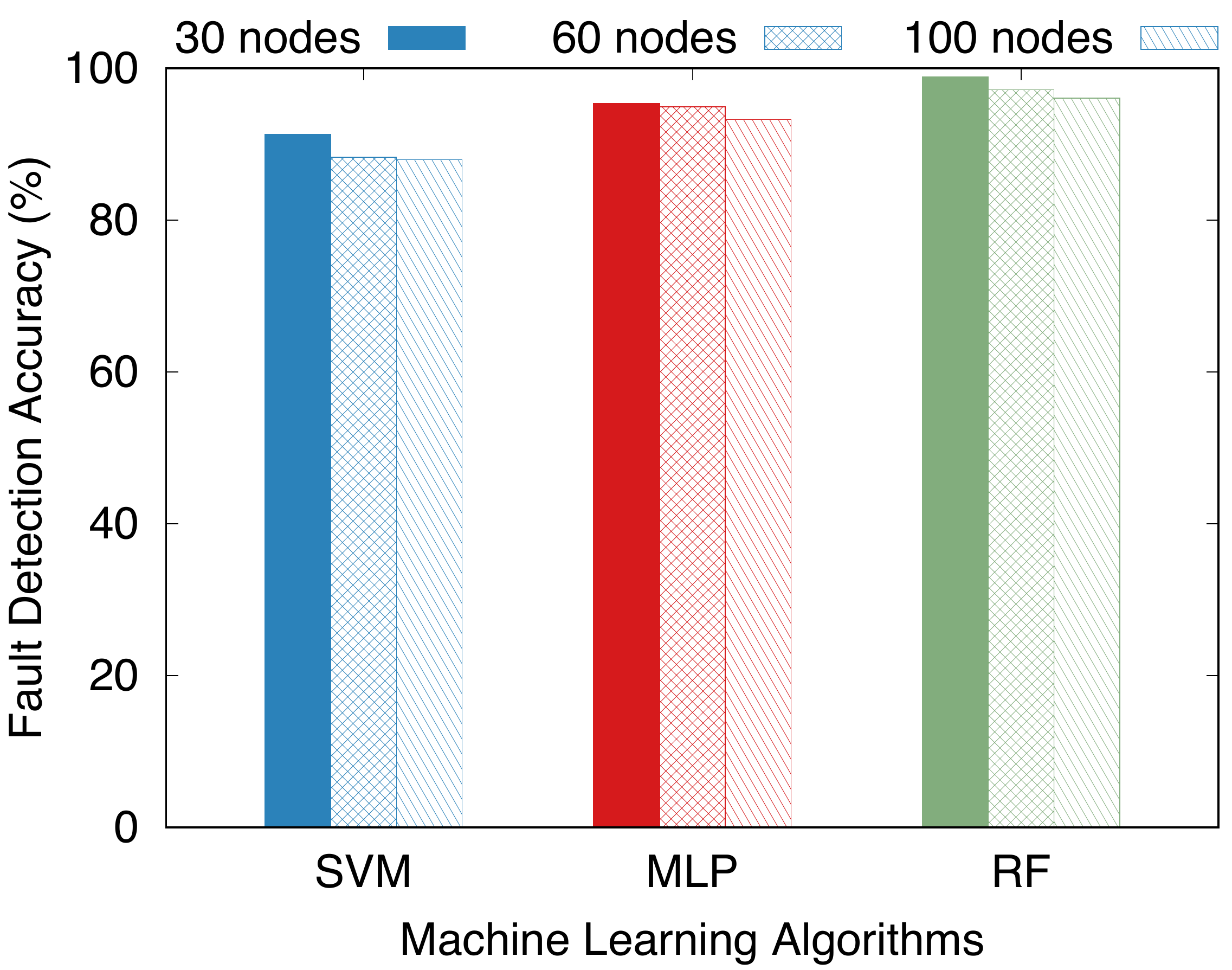}
  \caption{Fault detection accuracy of ML-LFIL.}
  \label{fig:fail_det}
\vspace{-1.5ex}
\end{figure}

As discussed earlier, the first stage of ML-LFIL identifies whether or not a link disconnection has occurred, using a link disconnection classifier. Instead of resulting in a ``no link fault'' message, the tentative disconnected link, $L_1$, (regardless of its correctness) will trigger the execution of the two subsequent stages for further analysis to identify and localize the link fault. It also should not be too sensitive since many false alarms could occur and unnecessarily trigger the analysis. We define the fault detection accuracy of ML-LFIL as the ratio of the number of data points associated with link faults that have been detected and triggered the execution of the second and third stages of ML-LFIL over the total number of data points associated with link faults that have actually occurred. In Fig.~\ref{fig:fail_det}, we present the fault detection accuracy of ML-LFIL for the three networks. It can be seen that the RF algorithm performs better than the other algorithms with fault detection accuracy of about $98.9\%$, $97.2\%$ and $96.1\%$ for the $30$-node network, $60$-node network and the 100-node Interroute network, respectively. This high fault detection accuracy results in the superior performance in identifying and localizing link fault in the two subsequent stages of ML-LFIL.
 
\subsubsection{Performance comparison with ping-based active probing}  
\label{subsec:Performance comparison}
We note that the recent literature considers only link disconnections. Thus, we only compare the performance of the first stage of the proposed method (ML-LFIL-S1) with the existing work~\cite{cheng:2016}. In~\cite{cheng:2016}, upon a link fault detection, the fault is localized by realizing two stages: (i) pinging all the nodes in the network to obtain sufficient data, and (ii) analyzing the obtained data. In Table~\ref{table:perf_comp}, we present the accuracy of ML-LFIL-S1 and the ping-based approach for different network topologies in localizing link disconnections in the network. We can see that for the 30-node network the ping-based approach and our ML-LFIL-S1 perform comparably, while for the 60-node and 100-node Interroute network, the ping-based approach performs slightly better than our ML-LFIL-S1. However, the ping-based approach requires significantly longer time than the proposed method to localize a link fault as discussed below.

\begin{table}[t]
\centering
\caption{Performance comparison (in \%)}
\begin{tabular}{ |@{\hspace{1ex}}c@{\hspace{1ex}}| l | @{\hspace{1ex}}c@{\hspace{1ex}} | @{\hspace{1ex}}c@{\hspace{1ex}} | @{\hspace{1ex}}c@{\hspace{1ex}} |} 
\hline
{\bf Topo.} & {\bf Algorithms} & {\bf Precision} & {\bf Recall} & {\bf $F_1$-Score}  \\ 
  \hline
  \hline
\multirow{2}{*}{$30$-node} &ML-LFIL-S1 & 98.63 & 98.14 & 98.38 \\
\cline{2-5}
                   &Ping-based & \multirow{2}{*}{99.45} & \multirow{2}{*}{99.12} & \multirow{2}{*}{99.28}\\
                    & approach & & & \\
\hline
\multirow{2}{*}{$60$-node} &ML-LFIL-S1 & 97.71 & 95.89 & 96.79\\
\cline{2-5}
                   &Ping-based & \multirow{2}{*}{99.38} & \multirow{2}{*}{98.91} & \multirow{2}{*}{99.14}\\
                     & approach & & & \\
\hline
\multirow{2}{*}{$100$-node} & ML-LFIL-S1 & 96.38 & 95.37 & 95.82\\
\cline{2-5}
                   &Ping-based &\multirow{2}{*}{99.12} & \multirow{2}{*}{98.50} & \multirow{2}{*}{98.81}\\
                     & approach &  & & \\    
\hline
\end{tabular}
\label{table:perf_comp}
\vspace{-2ex}
\end{table}  

\subsubsection{Fault Localization Time}  
\label{subsec:Fault Localization Time}
The time taken to localize a link fault in the network upon its occurrence is an important metric as this affects the failure recovery time. The faster the link fault localization, the less the impact of the fault on the network. In Table~\ref{table:fail_tim}, we present the time taken by ML-LFIL with different machine learning algorithms to localize a link fault in the network. Given a data point, the time taken by ML-LFIL is computed as the time to run all three stages to localize the disconnected and reconnected links. We compare the fault localization time with that of the ping-based active probing approach. Upon a link fault, we measure the time taken by the ping-based active probing approach to: (i) ping all the nodes in the network to obtain sufficient data, and (ii) analyze the obtained data to localize the fault. We note that the time taken to ping all the nodes depends on the propagation delay of the signalling messages on the links.  In our experiments with two random complex networks, we consider short connections where the propagation delay of links is randomly chosen in the range $[0.1,0.5]$ $\mu$s. This corresponds to the distance between nodes being in the range $[20,100]$ meters. For the 100-node network topology, we set the propagation delay based on the actual distance among nodes. The results show that ML-LFIL can localize a link fault in the order of microseconds. The worst case time of ML-LFIL when using MLP algorithm is $302.73\mu$s, whereas the ping-based approach requires significantly longer time to localize a link fault. It is worth mentioning that the fault localization time incurred by ML-LFIL does not vary much with the size of networks, i.e., increase in the size of feature vector evaluated by ML-LFIL. Whereas, the ping-based approach incurs increased localization time from $2.9$ms for the $30$-node network to $8.2$ms for the $60$-node network and $57$s for the $100$-node. It is to be also noted that the fault localization time of the ping-based approach will increase with the increasing link lengths. This shows that ML-LFIL enables a fast link fault identification and localization. 
 
\begin{table}[t]    
\centering
\caption{Comparison of fault localization time}
\begin{tabular}{ | l | r |  } 
\hline
\textbf {Methods} & \textbf{Time (in $\mu$s)}\\ 
\hline
    ML-LFIL with SVM  & $178.02$\\
\hline
    ML-LFIL with  MLP & $302.73$\\
\hline
    ML-LFIL with RF & $286.41$\\
\hline
Ping-based approach ($30$-node network) & $2960.24$\\
\hline
Ping-based approach  ($60$-node network)  & $8266.40$\\
\hline

Ping-based approach  ($100$-node network) & $56776575.82$\\
\hline
\end{tabular}
\label{table:fail_tim}
\vspace{-2.5ex}
\end{table}  

\section{Conclusion}  
\label{sec:conclusion}

In this paper, we developed a three-stage machine learning-based technique for link fault identification and localization (ML-LFIL) in complex networks. ML-LFIL learns the traffic behavior from the measurements captured from the network in normal working conditions and different fault scenarios that include link disconnections and link reconnections. The learning model takes into account the aggregate flow rate, end-to-end delay and packet loss captured at ingress and egress nodes. We trained ML-LFIL using different learning algorithms that include SVM, MLP and RF. We carried out comprehensive experiments in Mininet platform with two small-world complex networks and the 100-node Interroute network from the Internet topology zoo to study the performance of ML-LFIL. The results show that ML-LFIL achieves high performance in identification and localization of link faults with up to $97\%$ of accuracy. We compare ML-LFIL with a ping-based active probing approach. The results show that ML-LFIL requires significantly shorter time compared to the ping-based approach to achieve similar accuracy in link fault localization.    
  
  

\bibliographystyle{IEEEtran}  
\bibliography{ieeeiot2018}  

\newpage 
\vspace{2ex}
\begin{IEEEbiography}[{\includegraphics[width=1in,height=1.25in,clip,keepaspectratio]{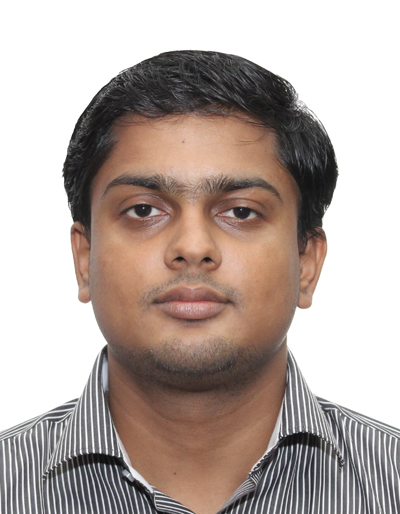}}]{Srinikethan Madapuzi Srinivasan} received the M.Sc. degree in Electrical and Computer Engineering with specialization in Communications and Networks from the National University of Singapore (NUS) in June 2016. He is currently working as a Research Engineer in the Communication and Networks Lab with the Department of Electrical and Computer Engineering at NUS since July 2016. He has been working to provide network services with improved QoS using Machine learning and Big Data. His research interests include software-defined networks, machine learning and Big Data.
\end{IEEEbiography}
  
\vspace{-6ex}
\begin{IEEEbiography}[{\includegraphics[width=1in,height=1.25in,clip,keepaspectratio]{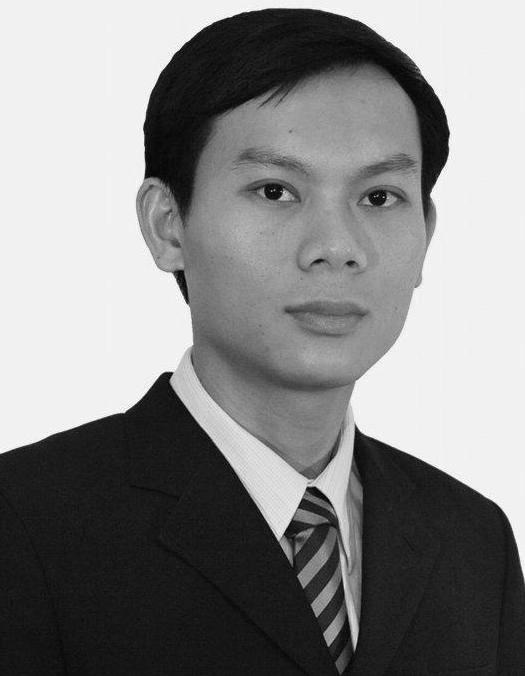}}]{Tram Truong-Huu} (M'12--SM'15) received the Ph.D. degree in computer science from the University of Nice Sophia Antipolis, France in 2010. He held a Post-Doctoral Fellowship with the French National Center for Scientific Research, from January 2011 to June 2012. In July 2012, he joined the National University of Singapore, where he is currently a Senior Research Fellow with the Department of Computer Science, School of Computing. His current research interests include scientific workflows, grid, cloud and mobile computing, software-defined networks, and Internet of Things. He received the Best Presentation Recognition at the IEEE/ACM UCC 2013. 
\end{IEEEbiography}

\vspace{-6ex}
\begin{IEEEbiography}[{\includegraphics[width=1in,height=1.25in,clip,keepaspectratio]{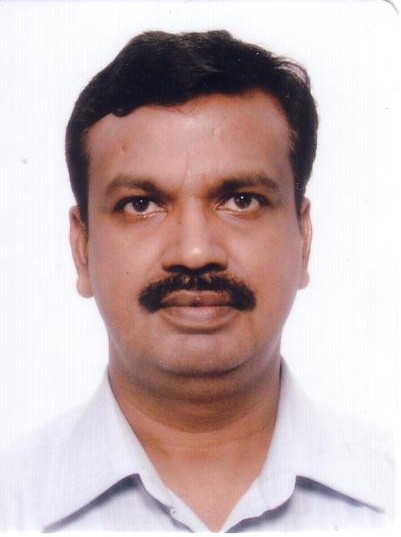}}]{Mohan Gurusamy}(M'00--SM'07) received the
  Ph.D. degree in Computer Science and Engineering from the Indian
  Institute of Technology, Madras in 2000. He joined the National
  University of Singapore in June 2000, where he is currently an
  Associate Professor in the Department of Electrical and Computer
  Engineering. His research interests are in the areas of Software Defined Networks, Network Function Virtualization, Internet of Things, Cloud Data Center networks, and Optical Networks. He has over 200 publications to his credit including two books and three book chapters in the area of optical networks. He is currently serving as an editor for IEEE Transactions on Cloud Computing, Elsevier Computer Networks journal and Springer Photonic Network Communications journal. He is serving (or) served as a TPC Co-Chair for several conferences including IEEE GLOBECOM 2019 (ONS) and IEEE ICC 2008 (ONS).
\end{IEEEbiography}

\end{document}